\documentclass[journal]{IEEEtran}
\usepackage{cite}

\ifCLASSINFOpdf 
\usepackage[pdftex]{graphicx}
\else
\usepackage[dvips]{graphicx}
\fi

\usepackage[cmex10]{amsmath}
\usepackage{algorithmic}
\usepackage[ruled,norelsize]{algorithm2e}

\usepackage{array}

\usepackage{mdwmath}
\usepackage{mdwtab}
\usepackage{todonotes}

\usepackage{eqparbox}

\ifCLASSOPTIONcompsoc
\usepackage[caption=false,font=normalsize,labelfont=sf,textfont=sf]{subfig}
\else
\usepackage[caption=false,font=footnotesize]{subfig}
\fi

\usepackage{fixltx2e}
\usepackage{url}
\usepackage{graphicx}
\usepackage{amsfonts}
\usepackage{amssymb}
\usepackage{amsthm}
\usepackage{bm}
\usepackage[ruled]{algorithm2e}
\usepackage[utf8]{inputenc}
\usepackage{booktabs}
\usepackage{epstopdf}
\usepackage{tabularx}
\usepackage{cite}
\usepackage{flushend}
\usepackage{multirow}

\definecolor{maroon}{cmyk}{0,0.87,0.68,0.32}

\newcounter{tempEquationCounter} 
\newcounter{thisEquationNumber}

\begin{document}
%
\title{Toward Real-Time Wireless Control of Mobile Platforms for Future Industrial Systems}

\author{
		Adnan~Aijaz,
		Aleksandar~Stanoev
        and~Mahesh~Sooriyabandara
        \vspace{-0.5cm}
\thanks{The authors are with the Bristol Research and Innovation Laboratory, Toshiba Research Europe Ltd., Bristol, BS1 4ND, U.K. Contact e-mail: adnan.aijaz@toshiba-trel.com}}


%
\markboth{IEEE International Conference on Computer Communications (INFOCOM) 2019 -- Demo Paper}%
{Shell \MakeLowercase{\textit{et al.}}: Bare Demo of IEEEtran.cls for Journals}
%


\maketitle

\begin{abstract}
The use of mobile platforms (MPs) is particularly attractive for various industrial applications. This demonstration highlights the importance of remote control of MPs and shows its viability over a high-performance wireless solution designed for closed-loop control. Further, it shows  the viability of formation control of a network of MPs through a leader-follower approach underpinned by high-performance wireless. 
\end{abstract}

\section{Introduction}

\IEEEPARstart{R}{ecent} technological advances have led to the realization of mobile platforms (MPs) like automated guided vehicles (AGVs), aerial drones and mobile robots. These MPs would likely change the conventional role of industrial machines, ultimately paving the way toward \emph{beyond 4.0} industrial systems \cite{aijaz_PIEEE}.
The logistic sector is already employing AGVs and mobile robots, particularly in warehouses and container terminals. Current industrial applications are mostly focused on a single MP. The use of a network of MPs, acting in a collaborative manner, improves efficiency and performance of industrial operations. Moreover, it also creates new opportunities for automation processes targeting complex tasks.

State-of-the-art  MPs are autonomous in nature and use different guidance control techniques like wire-guidance, magnetic tape, laser and camera-based imaging \cite{agv}. Such techniques not only require sophisticated sensing and localization capabilities at MPs but also need modifications to logistic facilities. Remote control\footnote{Remote control of real and virtual objects and processes is also the vision of the emerging \emph{Tactile Internet}.} of MPs is a disruptive paradigm that offers a low-cost alternative while providing various advantages. It is viable even with MPs  equipped with no or minimal sensory capabilities. It offers higher flexibility as MPs can be deployed without any pre-configuration of the logistic facility. It also provides instant reconfigurability in case of layout changes to facilities.  

Remote control of a MP creates a \emph{closed-loop control} scenario wherein command and feedback signals are exchanged between an external controller and a MP over a wireless network.  Such closed-loop control demands connectivity with very highly reliability and very low latency \cite{aijaz_PIEEE}.  This is to ensure stability of the control loop for accurate  path tracking under the imperfections of wireless medium. Closed-loop control also exhibits a cyclic traffic pattern that requires high determinism. The connectivity requirements become more stringent when multiple MPs are executing a collaborative task as this entails tight synchronization among MPs while maintaining a certain formation \cite{aijaz_PIEEE}. 

To this end, the main focus of this (\emph{live}) demonstration is real-time wireless control of an industrial system comprising multiple MPs. It has two main objectives. First, it demonstrates the viability of real-time closed-loop control of MPs through a high-performance wireless solution, known as \textsf{GALLOP}\footnote{\underline{G}ener\underline{A}lized c\underline{L}osed-\underline{L}oop c\underline{O}ntrol of \underline{P}rocesses}, which has been developed in authors' previous work \cite{GALLOP_patent}. Second, it demonstrates the viability of collaborative robotics over high-performance wireless through a leader-follower formation control architecture comprising multiple MPs.  

%
%
%

\begin{figure}[htbp]
\centerline{\includegraphics[scale=0.32]{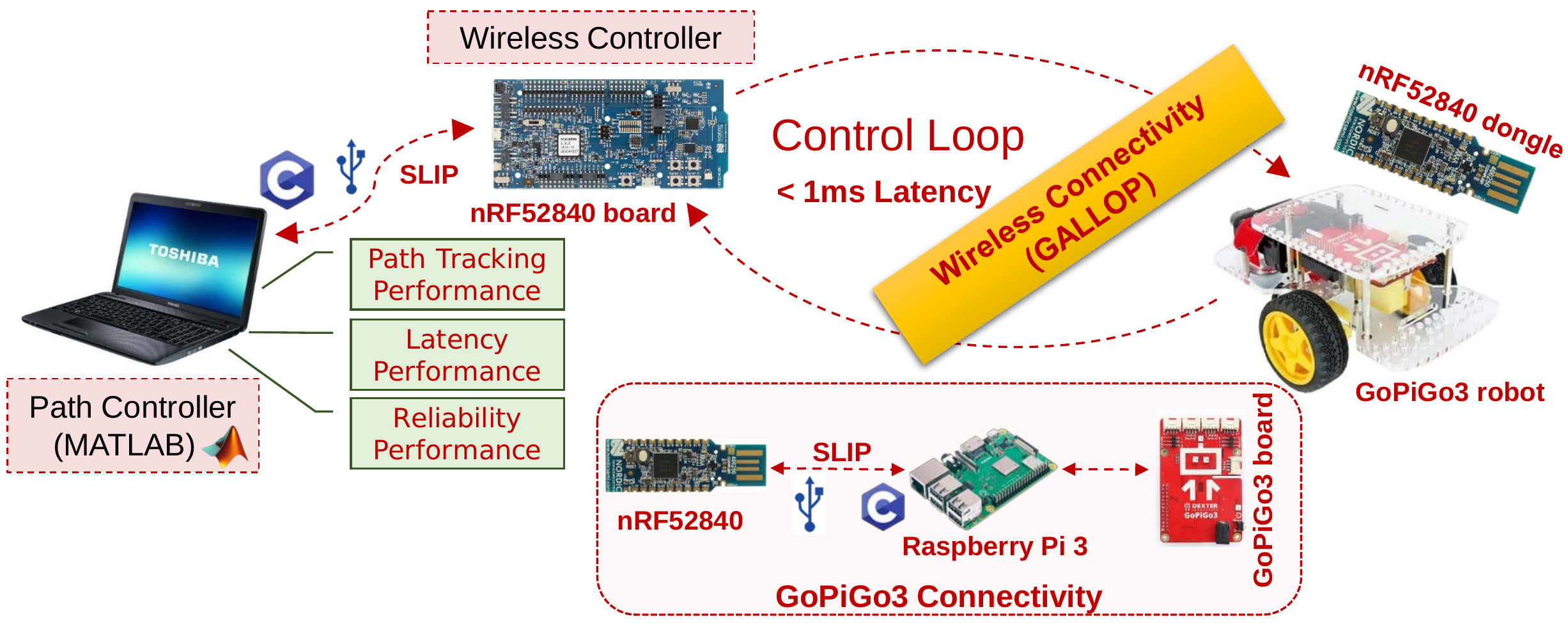}}
\caption{The remote control scenario and configuration.  }
\label{demo1}
\vspace{-1em}
\end{figure}

\section{Demonstration Overview}
\begin{figure}[htbp]
\centerline{\includegraphics[scale=0.31]{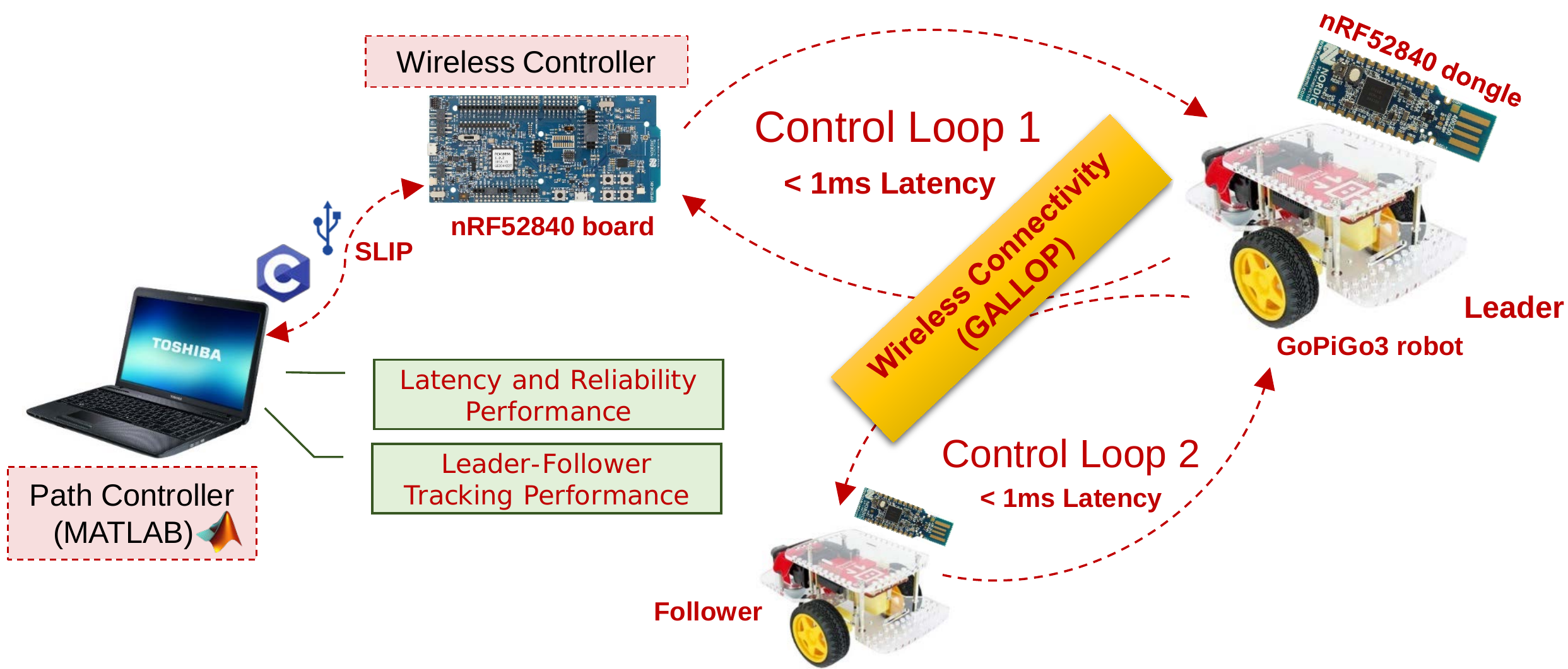}}
\caption{The leader-follower formation control scenario and configuration.  }
\label{demo2}
\vspace{-1em}
\end{figure}

The demonstration consists of two different scenarios. Both scenarios mimic an industrial environment wherein MPs are used for logistic applications e.g., a mobile robot or an AGV transporting goods in a warehouse or moving containers in a port terminal. The first scenario, which is shown in Fig. \ref{demo1}, demonstrates  wireless control of a mobile robot by an external controller which remotely drives it on a pre-defined path. In this case, the robot has no knowledge regarding the path. The second scenario (Fig. \ref{demo2}) demonstrates collaborative robotics wherein two mobile robots are required to maintain a certain formation, which is achieved through a leader-follower approach wherein the leader  is responsible for guiding the follower such that both maintain a desired formation. This results in a platooning scenario wherein the follower robot is remotely driven by the leader robot. In both scenarios, the control loops are closed over a wireless medium and create extremely stringent latency and reliability requirements for accurate real-time path tracking. 


\section{Design and Implementation}
\subsection{Mobile Robot}
We use the GoPiGo3 (\url{https://www.dexterindustries.com/gopigo3/}) robotic car which runs on a Raspberry Pi.
The GoPiGo3 robot is a differential drive system with two driving wheels (driven by motors having magnetic encoders) and one caster wheel. 
The motor control is performed by an ATMEGA328 microcontroller on the GoPiGo3 board which is stacked on top of the Raspberry Pi. The microcontroller  sends, receives and executes commands sent by the Raspberry Pi.

\subsection{Wireless System Design}
The proposed demonstration employs \textsf{GALLOP} as the underlying wireless technology. \textsf{GALLOP} has been specifically designed to provide high-performance connectivity for realizing closed-loop control. \textsf{GALLOP} is agnostic to the Physical (PHY) layer and can be implemented on any off-the-shelf wireless chipset.  The medium access control (MAC) layer is based on time division multiple access (TDMA), frequency division duplexing (FDD) and frequency hopping.  \textsf{GALLOP} caters for the peculiarities of closed-loop operation through a control-aware bi-directional distributed scheduling algorithm.  It  is capable of operating in both single-hop and multi-hop topologies. \textsf{GALLOP} implements novel cooperative diversity and efficient retransmission techniques for achieving very  high reliability. Detailed description of \textsf{GALLOP} is available in \cite{GALLOP_patent}.

We have implemented \textsf{GALLOP} on Nordic Semiconductor nRF52840 platform (\url{https://www.nordicsemi.com/eng/Products/nRF52840}) which is built around a 32-bit ARM Cortex-M4F CPU with 1 MB flash and 256 kB RAM on chip. The embedded 2.4 GHz transceiver supports multiple protocols including Bluetooth 5, Bluetooth low energy, and IEEE 802.15.4.  \textsf{GALLOP} has been implemented on the uncoded 2 Mbps Bluetooth 5 PHY layer. \textsf{GALLOP} adopts a flooding-based protocol \cite{glossy} for time synchronization. The transmit power is set to 8 dBm and the net MAC payload is 16 bytes. 
The wireless  controller has been implemented on the nRF52840 board whereas the GoPiGo3 robotic cars are equipped with the nRF52840 dongles. 

\subsection{Path Controller Design and Interfacing}
\begin{figure}[htbp]
\centerline{\includegraphics[scale=0.22]{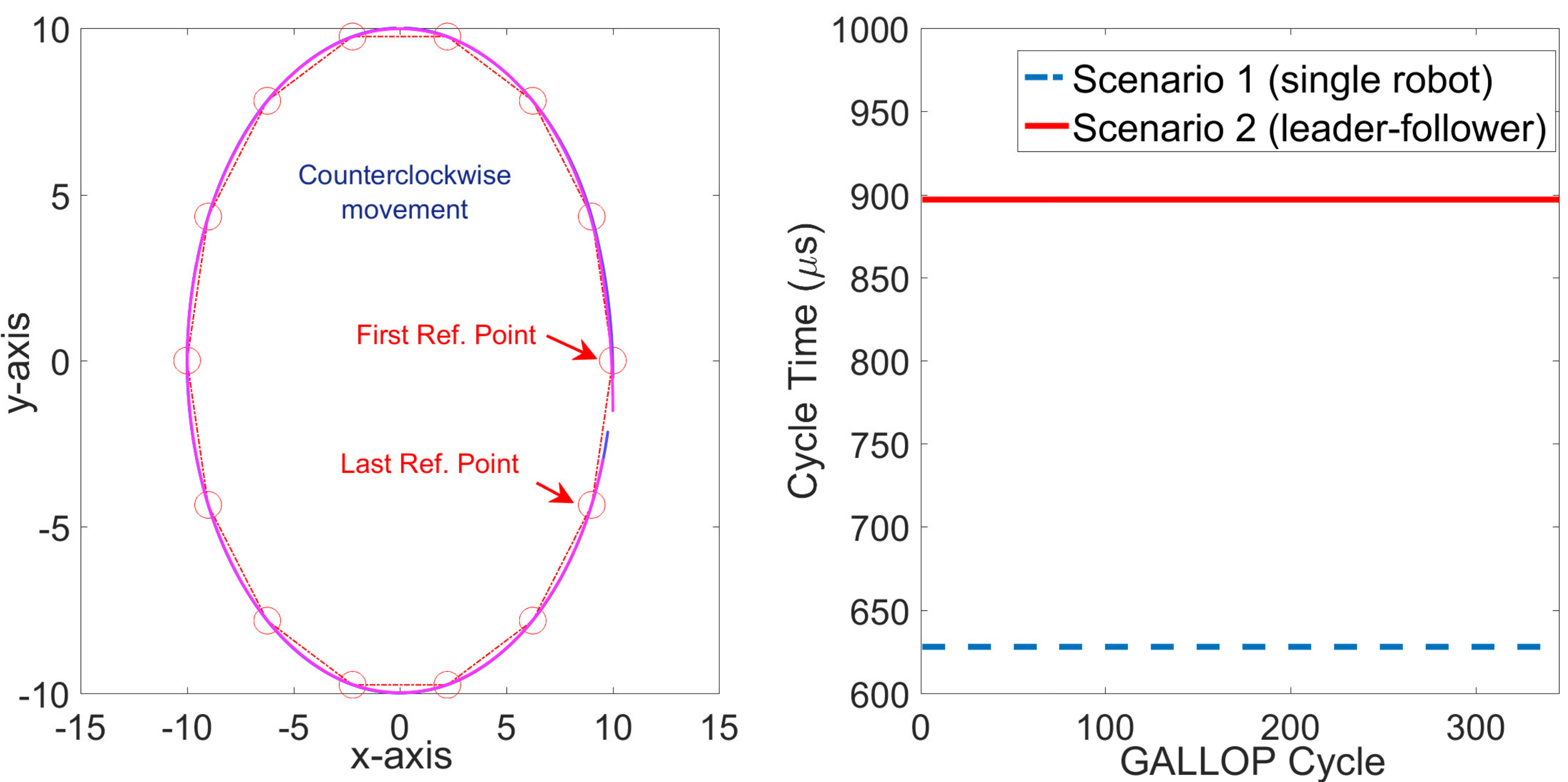}}
\caption{Performance evaluation: (a) leader-follower path tracking (small circles denote reference points of the path); (b) cycle time (latency) of \textsf{GALLOP}. }
\label{demo3}
\vspace{-1em}
\end{figure}

The objective of the path controller (PC) is to remotely drive the GoPiGo3 robot along a pre-defined route by controlling the speed of the driving wheels. The PC has route information as a set of reference points. It periodically receives motor encoder values for left and right wheel as feedback from the robot. It reads the first reference point and calculates the deviation error from the current position of the robot which is obtained using the feedback. It estimates the required speed of the wheels based on a quadratic curve approach and a kinematic model  characterizing the robot's movement \cite{bk_amr}.  This information is sent as the control command on the forward path. The algorithm continues until the robot reaches the reference point. The procedure is repeated for subsequent reference points until the robot reaches its destination. 
The path control functionality for platooning scenario follows a similar approach. However, in this case the PC logically resides in the leader GoPiGo3 robot and the set of reference points for the follower GoPiGo3 robot is populated on-the-fly using  leader's position. A few results from the demonstration are shown in Fig. \ref{demo3}.

The PC also handles an emergency stop function which shows  \textsf{GALLOP}'s ability to support timely delivery of event-driven information.  The GoPiGo3 robots are equipped with an infrared distance sensor and the distance feedback is piggybacked over the encoder feedback. Emergency stop functionality is triggered if either of the robots detect an obstacle on the route. On apprehending a collision, the PC sends a control message to stop the robotic system. 

The nRF52840 board communicates with a C application over a USB interface through a serial line Internet protocol (SLIP). The C application communicates with the PC which has been implemented in MATLAB. On the GoPiGo3 car, the nRF52840 dongle communicates with Raspberry Pi over the aforementioned USB interface.


\section{Remarks}
This demonstration shows the effectiveness of \textsf{GALLOP} in handling different closed-loop control scenarios. Vertical applications of \textsf{GALLOP} and the proposed system design  include formation control of an aerial drone fleet, nuclear decommissioning, highway platooning, virtually coupled train systems, and collaborative operation of multiple robots. 
A short \textbf{video} of the demonstration is available at [\url{https://www.dropbox.com/s/roctb3pac5o8y53/GALLOP_demo_final.mp4?dl=0}].







\bibliographystyle{IEEEtran}

\bibliography{IEEEabrv,JSAC_bib}
%

\end{document}